\documentclass[11pt]{pazh}

\usepackage{psfig}

\begin{document}
	
\setcounter{page}{283}

\sloppypar

\title{\bf A photometric study of faint galaxies in the field
of GRB\,000926}

\author{\copyright\,2004\,\,\, T.A.Fatkhullin$^1$, A.A.Vasil'ev$^2$, 
V.P.Reshetnikov$^2$} 

\institute{Special Astrophysical Observatory, Russian Academy of
Sciences, Nizhniii Arkhyz, 357147 Karachai-Cherkessian Republic, Russia
\and 
Astronomical Institute of St.Petersburg State University,
Universitetskii pr. 28, Petrodvoretz, 198504 Russia}

\authorrunning{Fatkhullin et al.}
\titlerunning{Faint galaxies}

\abstract{We present our $B$, $V$, $R_c$, and $I_c$ observations of a 
3.6$'$$\times$3$'$ field centered on the host galaxy of GRB\,000926
($\alpha_{2000.0}$=17$^h$04$^m$11$^s$, 
$\delta_{2000.0}$=+51$^{\rm o}$47$'$9.8$''$). The observations were carried 
out on the 6-m Special Astrophysical Observatory telescope using the 
SCORPIO instrument. The catalog of galaxies detected in this field includes 
264 objects for which the signal-to-noise ratio is larger than 5 in
each photometric band. The following limiting magnitudes in the catalog 
correspond to this limitation: 26.6($B$), 25.7($V$), 25.8($R_c$), and 
24.5($I_c$). The differential galaxy counts are in good agreement with
previously published CCD observations of deep fields. We estimated the 
photometric redshifts for all of the cataloged objects and studied the 
color variations of the galaxies with $z$. For luminous spiral galaxies
with $M(B) < -18$, we found no evidence for any noticeable evolution of 
their linear sizes to z $\sim$1.
}
\titlerunning{Faint galaxies}
\maketitle

\section{Introduction}

In recent years, significant progress has been made in studying the 
properties and evolution of distant galaxies. An important role in this 
progress has been played by detailed studies of several so-called
deep fields -- relatively small areas imaged (generally in several color 
bands) with long exposure times. The best-known deep fields include 
the northern and southern fields of the Hubble Space Telescope 
(Ferguson et al. 2000), Subaru (Maihara et al. 2001), VLT FORS 
(Heidt et al. 2003), and the most recent and deepest (in the history of 
optical astronomy) survey conducted as part of the GOODS (Great
Observatories Origins Deep Survey) project with the Hubble Space Telescope 
(Giavalisco et al. 2003). In addition to surveys covering an appreciable 
part of the sky (2MASS, 2dF, SDSS, and others), but limited to relatively 
low redshifts ($z\leq0.3$), deep fields allow the properties of high $z$ 
galaxies to be studied and provide information about the evolution of the
integrated parameters of galaxies.

Our primary objective was to study in detail the faint galaxies distinguished 
in the deep field obtained with the 6-m Special Astrophysical Observatory
(SAO) telescope as part of our program of optical identification of 
gamma-ray bursts (GRBs). In recent years, investigators have increasingly 
pointed out that, with the accumulation of observational data,
GRBs with their afterglows and host galaxies are becoming a useful tool 
in observational cosmology (see, e.g., Djorgovski et al. 2003; Ramirez-Ruiz
et al. 2001; Trentham et al. 2002; and references therein). Thus, a 
comparative analysis of the properties of GRB host galaxies with the 
properties of galaxies at the same redshifts is now of relevant
interest. One of the methods for solving this problem is to study the 
population of faint galaxies in the deep fields of GRB host galaxies.

\section{Observations and data reduction}

We carried out our photometric observations of the field of the host galaxy 
of GRB\,000926 on July 24 and 25, 2001, using the 6-m SAO telescope. 
The observing conditions were photometric with 1.3-arcsec seeing, 
measured as the full width at half maximum (FWHM) of the images of
starlike objects in the field. The field was centered on the coordinates 
of the host galaxy $\alpha_{2000.0}$=17$^h$04$^m$11$^s$, 
$\delta_{2000.0}$=+51$^{\rm o}$47$'$9.8$''$, which correspond to the 
Galactic latitude and longitude $b =37^{\rm o}21'$ and $l = 77^{\rm o}45'$ , 
respectively. According to the infrared maps taken from the paper by 
Schlegel et al. (1998), the Galactic reddening toward the field
being studied is $E(B-V)$=0.023. 

In our observations, we used the SCORPIO (Spectral Camera with Optical 
Reducer for Photometrical and Interferometrical Observations; for a 
description, see http://www.sao.ru/moisav/scorpio/scorpio.html) 
instrument mounted at the prime focus of the 6-m SAO telescope. 
The detector was a TK1024 1024$\times$1024 CCD array. The pixel size 
was 24 $\mu$m, which corresponded to an angular scale of 0.289$''$ per pixel.
The CCD response curve and the SCORPIO broad-band filters reproduced 
a photometric system close to the standard Johnson--Cousins $BVR_cI_c$ 
system (Bessel 1990). We took five frames in each of the $B$ and $V$ 
bands with total exposure times of 2500$^s$ and 1500$^s$, respectively; 
ten frames in $R_c$ (1800$^s$); and fifteen frames in $I_c$ (1800$^s$). 
The absolute photometric calibration of the data was performed
using the observations of standard stars from the lists by Landolt (1992) 
and Stetson (accessible at http://cadcwww.dao.nrc.ca/cadcbin/wdb/astro-
cat/stetson/query) on the same nights. 

\begin{figure*}[!ht]
\centerline{\psfig{file=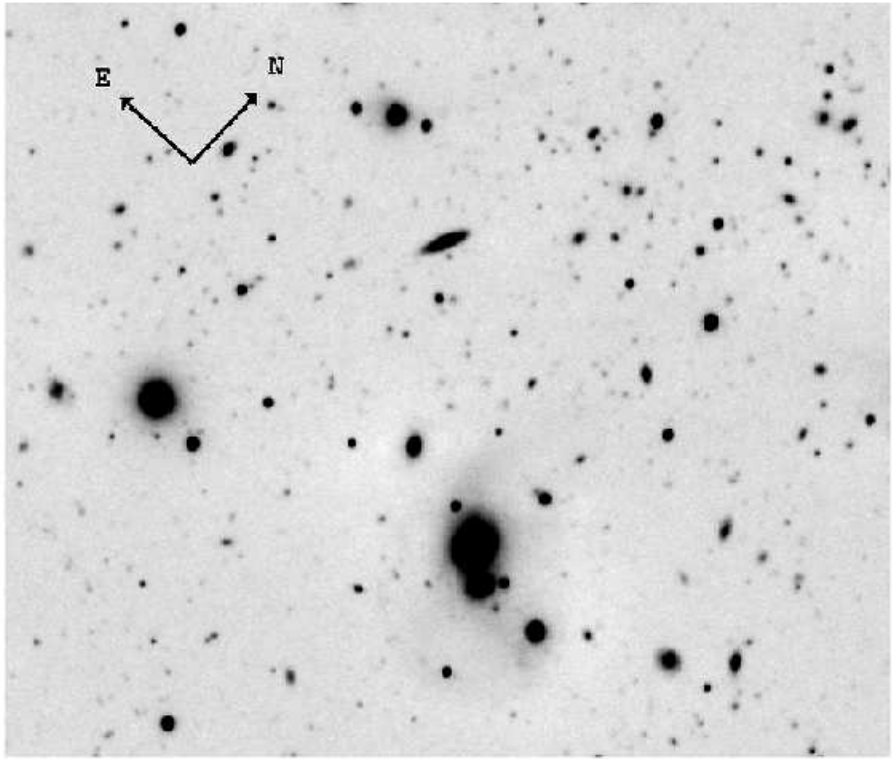,width=12cm,clip=}}
\caption{Reproduction of the field of GRB\,000926. The field
sizes are 3.6$'$$\times$3$'$.}
\end{figure*}

The primary data reduction was performed using the 
ESO-MIDAS\footnote{This package is supported and distributed by the European
Southern Observatory.} software package. It included debiasing, flat 
fielding, defringing in $R_c$ and $I_c$, and cosmic-ray hit removal 
(for a detailed description, see the dissertation by Fatkhullin 2003). 
All of the frames taken in one color band were coadded. The coadded frames 
were reduced to the same orientation and to a single coordinate system. 
The sizes of the region where the coadded images overlapped in all
bands were 3.6$'$$\times$3$'$. This region was used for the
subsequent analysis (Fig. 1).

\section{Large-scale photometry of objects in the field} 

\subsection{Extracting objects}

We used the SExtractor (Source Extractor) (Bertin and Arnouts 1996) package 
to extract objects in the field and to perform their photometry. This package
is unique in that it allows us to extract objects and construct their 
isophotes on one image and to calculate the fluxes within these isophotes 
on another image. Thus, we can determine the fluxes in all bands
for a given object within the same region of the image. On the other hand, 
this technique makes it possible to find all of the objects in the field 
from its common image for all bands, which allows problems with the
identification of the same object in different bands to be avoided.

There are several methods for constructing a common image (i.e., the 
detection field). For example, the field obtained by coadding the frames 
in different color bands is often used to extract objects. We used a
relatively new method, that of constructing a $\chi^2$-field.
The $\chi^2$-field is a common image for all bands, and it is used to 
extract objects and to construct a common catalog of objects for all bands. 
The idea of the method is that probabilistic methods are used to analyze the
distribution of image pixel values and to determine the optimum detection 
limit for objects above the background level (Szalay et al. 1999).

Schematically, the process of constructing the $\chi^2$-field can be 
described as follows. The mean and rms deviation ($\sigma_b$) of the sky 
background (the distribution of sky background values is assumed
to be Gaussian) are determined from the images in each color band. 
Subsequently, the corresponding background value is subtracted from each 
image, and the resulting frame is divided by $\sigma_b$. As a result, we
obtain a transformed image for each band in which the sky background is 
specified by a Gaussian with a zero mean and a unit variance. If we 
consider the combined transformed images, each pixel of the common field
may be represented as a vector with the number of elements equal to the 
number of bands. The resulting $\chi^2$-field is an image, each pixel of 
which is equal to the square of the length of the corresponding vector
of the common field.

If the observed region contained no objects and displayed only the sky 
background, then the probability distribution function for $\chi^2$-field 
values would correspond to $\chi^2$ with the number of degrees of freedom
equal to the number of bands. However, an actual field contains objects, 
and, therefore, the distribution of $\chi^2$-field values is distorted. 
By analyzing the $\chi^2$-fields for the observed region and for the sky 
background, we can determine the optimum detection limit for objects in 
the actual field; this limit turns out to be lower than that in the method 
of coadding the frames in different color bands.

The detection conditions were specified in such a way that objects that 
occupied an area of no less than five pixels were extracted at fluxes in 
these pixels exceeding 3$\sigma$ above the background level. For detection, 
we also used filtering based on a wavelet analysis using a mexhat-type 
filter with a FWHM that corresponded to the quality of our images, i.e.,
1.3$''$. This filtering allowed us to effectively extract objects in frames 
with a large image density and around bright extended galaxies as well as 
to separate the components of multiple objects.

\subsection{Object extraction results and photometry}

More than 550 objects satisfied the detection criteria. However, to 
eliminate problems related to inaccurate photometry due to the defects of 
the objects themselves, we excluded those which turned out to be
near the image boundaries and in two regions around very bright objects 
(an overexposed star and a large interacting system of galaxies). Thus, 
slightly more than 400 objects remained.

To increase the reliability of the subsequent analysis, we included in 
the final catalog only those objects for which the signal-to-noise ratio 
was larger than 5 in each band. There were 285 such objects.
The following limiting magnitudes in the catalog corresponded to this 
limitation: 26.6 ($B$), 25.7 ($V$), 25.8 ($R_c$), and 24.5 ($I_c$).

For each object, we determined the total apparent magnitudes and the 
corresponding color indices using the so-called best magnitudes generated by
the SExtractor package. These values are the best fits to the asymptotic 
magnitudes of extended objects (Bertin and Arnouts 1996).

The SExtractor package allows the objects to be separated into starlike and 
extended ones by assigning a corresponding ``stellarity index'' from 0 
(extended) to 1 (star) to each of them. For the subsequent analysis, we 
selected only those objects for which the stellarity index was less 
than 0.7 in $B$ (in this band, the stellar disks and spiral arms of galaxies
are separated relatively better). There were 264 such objects. We attributed 
the objects with an index larger than 0.7 (their number was 21) to stars.

\section{Estimating the redshifts}

Determining the spectroscopic redshifts for several hundred faint objects 
extracted in deep fields is a complicated and time-consuming observational
problem. Fortunately, for many problems (e.g., for estimating the evolution 
of the galaxy luminosity function), the so-called photometric redshifts 
estimated from multicolor photometry turn out to be quite acceptable. 
The accuracy of these $z$ estimates is about 10\%, which is high enough 
for statistical studies of the properties of distant objects. The main
idea of the photometric redshift estimation is very simple: an object's 
multicolor photometry may be considered as a low-resolution spectrum that 
is used to estimate $z$ (Baum 1963).

In practice, we estimated the photometric redshifts for the extended 
objects of our sample using the Hyperz software package 
(Bolzonella et al. 2000). The input data for Hyperz were: the apparent 
magnitudes of the objects in four bands, the internal extinction law
(we used the law by Calzetti et al. (2000) for starburst galaxies, which 
is most commonly used for studies similar to our own), the redshift range 
in which the solution is sought (we considered $z$ from 0 to 1.3), and
the cosmological model (we used a flat model with 
$\Omega_m$=0.3, $\Omega_{\Lambda}$=0.7, and $H_0$= 70 km/s/Mpc).
The observed spectra of the galaxies were compared with the synthetic 
spectra for E, Sa, Sc, Im, and starburst galaxies taken from the GISSEL98 
library.

Apart from the redshift estimate, the Hyperz package yielded the following 
basic parameters for each object: the spectral type of the galaxy (based 
on the similarity between the spectral energy distribution of the object 
and one of the synthetic spectra), the age, the absolute $B$ magnitude, 
and a number of others.

\section{Results and discussion}

Our catalog of extended objects ($N$=264) has the following integrated 
parameters:

(1) the mean absolute magnitude of the galaxies is
$M_B$ = --17.4$\pm$2.7 ($\sigma$);

(2) the mean photometric redshift is $\langle z \rangle$ = 0.53$\pm$0.39;

(3) the stellarity index in the $B$ band is 0.08$\pm$0.12, i.e., extended 
objects dominate in the catalog;

(4) the apparent flattening of the galaxies varies between 0.22 and 1.00 
with a mean value of $\langle b/a \rangle$ = 0.80$\pm$0.20.

Irregular and starburst galaxies (126 of the 264 objects or 48\%) constitute 
about one half of the catalog. Sa--Sc spiral galaxies are encountered in 
almost a third of the cases (79 objects or 30\%). The contribution of 
elliptical galaxies is 22\% (59 objects).

The mean angular size of the objects (FWHM) corrected for the FWHM of 
the stars varies between 1.0$''$ in $R_c$ to 1.2$''$ in $I_c$ (the errors 
of the means are $\pm$0.5$''$). At the mean redshift, a linear size of  
$\approx$7 kpc corresponds to this angular size. If we restrict our
analysis to relatively near galaxies at $z\leq0.5$ (in this case, the 
mean redshifts of the galaxies of various spectral types are close), 
the expected dependence of the linear sizes on the galaxy type shows up. 
Thus, the mean FWHM for elliptical galaxies corresponds to a linear size 
of 3.4 kpc, while starburst galaxies exhibit more extended brightness 
distribution with FWHM $\approx$4.7 kpc.

\subsection{Starlike objects}

We used a fairly stringent criterion to separate galaxies and stars: we 
attributed objects with a stellarity index in $B$ of less than 0.7 to 
galaxies (a less stringent limitation, e.g., 0.9--0.95 (Arnouts et al. 2001), 
is commonly used).

The number of stars found in the field (21 stars with apparent magnitudes 
$B\leq24.5$) is in good agreement with the prediction of Bezanson's Milky
Way model (for a description of this model, see, e.g., Robin et al. 2000). 
In a field with an area of 0.003 square degrees located at the Galactic
coordinates that correspond to our region, this model predicts 24 stars 
with $B\leq24.5$. Given the small size of the field under study, such a 
close coincidence is, of course, partly accidental. However, it is indicative
of the relatively small number of stars that could be erroneously attributed 
to galaxies.

Figure 2 shows the distribution of stars in the two-color diagram. The 
solid line in the figure represents the color sequence for main sequence 
stars (Sparke and Gallagher 2000). Most of the stars exhibit colors
typical of main-sequence dwarfs.

\begin{figure}[!ht]
\centerline{\psfig{file=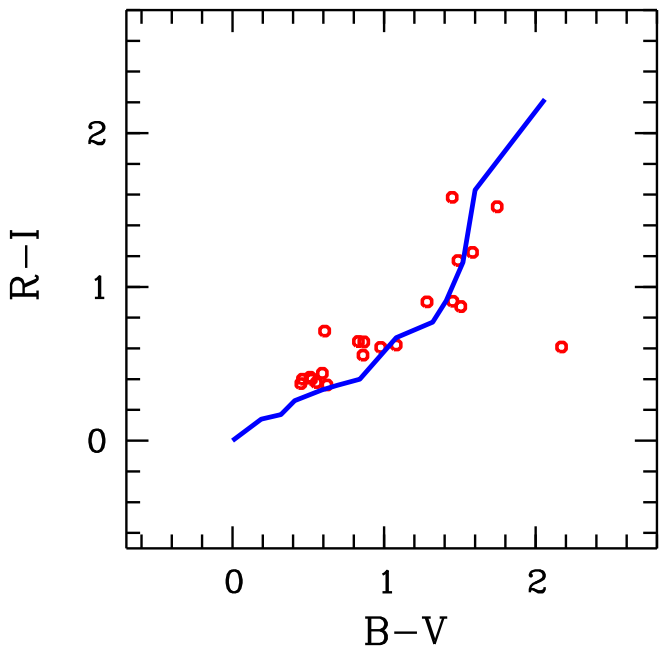,width=9cm,angle=-90,clip=}}
\caption{Color distribution for starlike objects in the two-color
diagram. The solid line represents the color sequence for main sequence
stars.}
\end{figure}

\subsection{Galaxy counts}

The differential galaxy counts normalized to 1 sq. degree and found 
within 0$^m$.5 bins are shown in Fig. 3 (circles). These counts were not 
corrected for observational selection and represent the actually observed 
numbers. The crosses in the figure indicate similar counts for the VIRMOS 
survey (Le Fevre 2003; McCracken et al. 2003). Clearly, our counts are in 
excellent agreement with this survey. For apparent magnitudes 
$B < 25$, $V < 24.5$, $R_c < 24$, and $I_c < 24$, the difference in the 
numbers of galaxies observed in one magnitude bin, on average,
does not exceed 20\%. For fainter magnitudes, observational selection, 
which manifests itself in the form of bends in the curves of differential 
counts in Fig. 3, strongly affects our data.

\begin{figure*}[!ht]
\centerline{\psfig{file=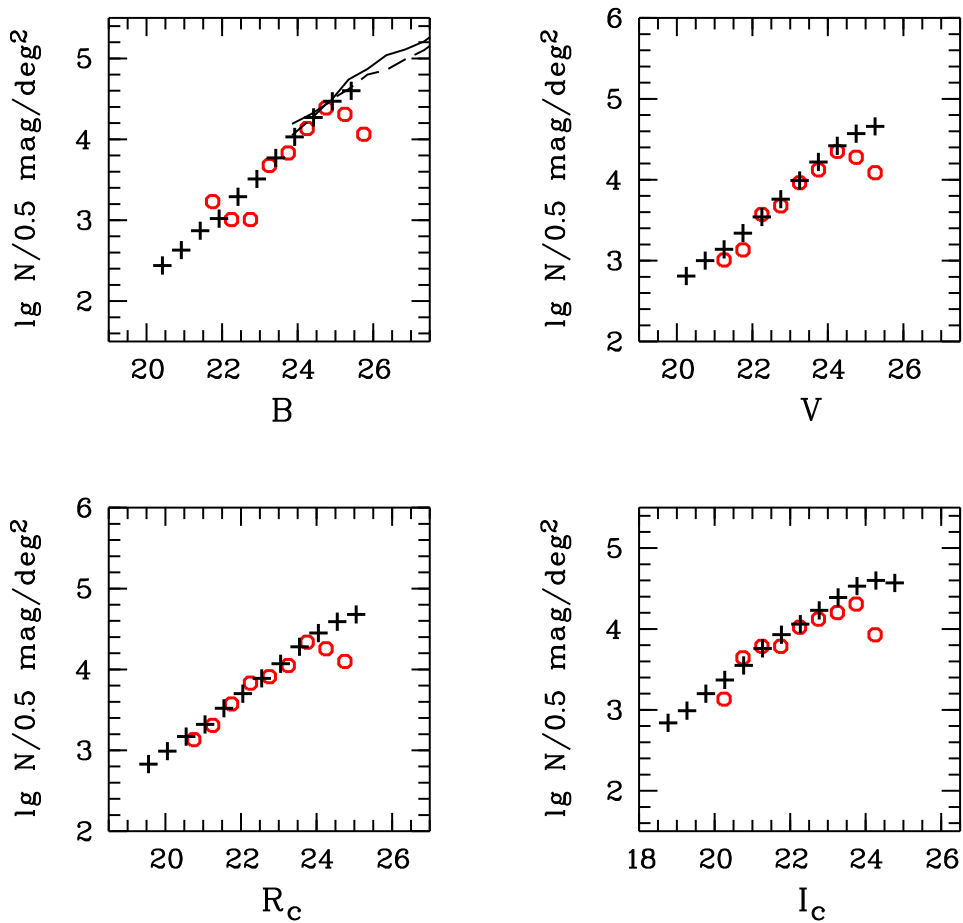,width=13cm,angle=-90,clip=}}
\caption{Differential galaxy counts in the field of GRB\,000926 (circles)
in different bands. The crosses indicate similar counts for
the VIRMOS survey (McCracken et al. 2003). The solid and dashed lines 
represent the $B$-band counts for the northern and southern deep fields 
of the Hubble Space Telescope, respectively (Metcalfe et al. 2001).}
\end{figure*}

The slopes of the curves shown in Fig. 3 are also in good agreement. 
According to our data, these slopes are \\
0.47 $\pm$ 0.09 ($B \leq 25$),  0.47 $\pm$ 0.04 ($V \leq 24$),\\
0.38 $\pm$ 0.03 ($R_c \leq 24$) and 0.29 $\pm$ 0.04 ($I_c \leq 24$).
A similar variation in the slopes of the curves of differential counts 
when passing from $B$ to $I_c$ was also pointed out by McCracken et al. 
(2003).

On the other hand, the VIRMOS data are in excellent agreement (to within 10\%) 
with previous CCD galaxy counts (McCracken et al. 2003). As an example, 
Fig. 3 shows the counts for the northern and southern deep fields of the 
Hubble Space Telescope (Metcalfe et al. 2001). Near $B\sim 25$, the ground-
based and space counts join to form a single dependence that is currently 
traceable to $B\sim 29$. A detailed interpretation of this dependence 
(apart from the fact that this is a classical cosmological test) can yield
important information about the evolution of galaxy properties (Gardner 1998).

In Fig. 4, the observed colors of our field galaxies are plotted against 
their apparent magnitude. The figure clearly shows a trend well known from 
previous studies: fainter galaxies have, on average, bluer colors. The 
dashed lines in the figure indicate similar dependences for the VIRMOS 
galaxies (McCracken et al. 2003) that are in satisfactory agreement with
our data.

\begin{figure}[!ht]
\centerline{\psfig{file=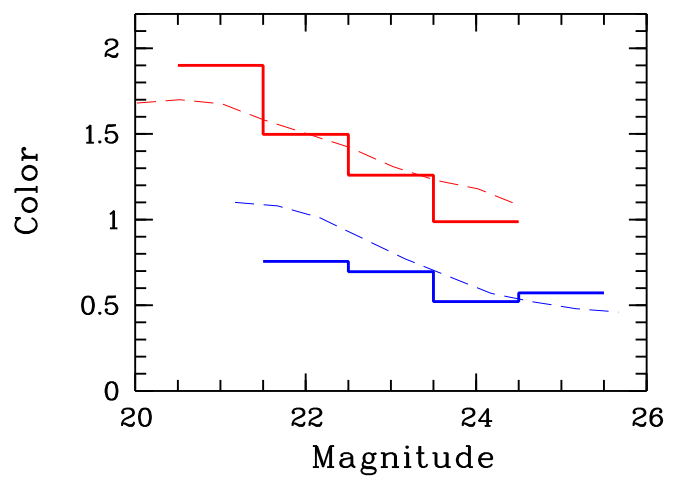,width=9cm,angle=-90,clip=}}
\caption{Mean color index $V-I_c$ for the galaxies of our field
versus apparent magnitude $I_c$ (red), and color index $B-V$
versus $B$ (blue). The dashed lines represent similar diagrams
for the VIRMOS galaxies (McCracken et al. 2003).}
\end{figure}

The agreement between the counts in our field with the data for other 
deep fields is remarkable, because our field is very small and contains 
a relatively small number of galaxies. For example, the previously
mentioned VIRMOS catalog contains $\approx10^5$ galaxies; this number 
exceeds the size of our sample by several hundred. This agreement suggests 
that our field is representative and reflects well the average
properties of the galactic population, at least up to $z\sim1$.

\subsection{The $z$ distribution}

Figure 5 (left) shows the redshift distribution of the catalogued galaxies. 
This distribution exhibits a peak at $z = 0.2-0.3$ and a long, gently 
sloping tail up to $z = 1.3$. The $z$ distribution of the galaxies is 
distorted by selection effects. One of the strongest selection
effects is our requirement that the signal-to-noise ratio for an object 
be larger than five in all four bands. As a result of this condition, 
early-type objects were mostly detected at a relatively low $z$ because of 
the decline in the spectral energy distribution of early-type galaxies
at short wavelengths. This effect is clearly seen from Fig. 5 (left), 
where the dashed and heavy solid lines indicate the distributions of E--Sa 
and Sc--Im galaxies, respectively.

Figure 5 (right) shows the distribution of the catalogued galaxies in the 
"absolute magnitude ($M(B)$) -- redshift" plane. The galaxy distribution 
in this plane is also determined by selection effects; we extract
mostly luminous objects among the more distant objects. As an example, 
the dashed line in the figure indicates the selection curve for an Sc spiral 
galaxy with the apparent magnitude $B = 26$ (the observed magnitude distorted 
by the $k$-correction). If all of the galaxies of this type brighter than 
26$^m$ were included in our sample of objects, they would lie below the
dashed line in the figure. The fainter galaxies with $B > 26$ would lie 
above this line. Clearly, the magnitude limitation explains well the 
galaxy distribution in the $M(B) - z$ plane.

\begin{figure*}[!ht]
\centerline{\psfig{file=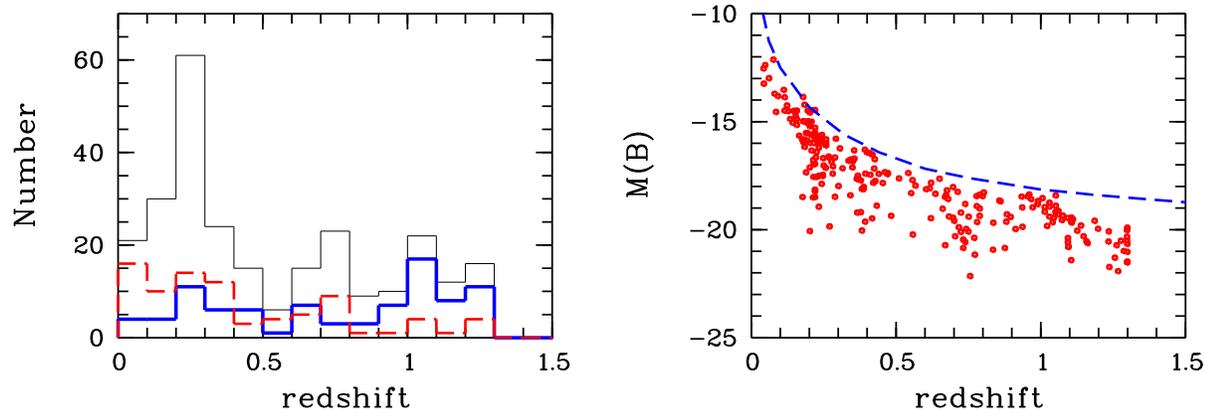,width=16cm,angle=-90,clip=}}
\caption{The $z$ distribution of sample galaxies (thin solid line); 
the heavy solid (blue) and dashed (red) lines indicate the distributions for
Sc--Im and E--Sa galaxies, respectively (left). The absolute 
magnitude -- redshift diagram (right).}
\end{figure*}

The strong selection in absolute magnitude in our sample makes it difficult 
to directly compare the parameters of nearby and distant galaxies. However,
if we restrict our analysis only to luminous objects (say, with $M(B) < -18$), 
then the selection effect is weaker for them, and such galaxies are 
confidently extracted up to the limiting $z$ of our sample (Fig. 5).
Spiral and starburst galaxies dominate (90\%) in this sample.

Figure 6 shows variations in the angular sizes (the FWHM corrected for the 
FWHM of the stars) of luminous ($M(B) < -18$) galaxies in the $R_c$ band (the
seeing was best during observations in this band). Note that the FWHM values 
depend on the cosmological deeming in brightness much less strongly
than do the isophotal diameters. Therefore, they may be used as rough 
estimates of the galaxy sizes. The barred circles in Fig. 6 represent the 
mean values and the corresponding dispersions for the redshift
ranges 0--0.5, 0.5--1.0, and 1.0--1.3, while the dashed lines indicate the 
lines of constant linear sizes (the lower and upper curves correspond to 
linear sizes of 3 and 15 kpc, respectively). The angular sizes of
the extended objects from our sample are located, on average, between the 
two curves of constant linear sizes. For the luminous galaxies under 
consideration, the mean FWHM in linear measure changes only slightly with 
increasing $z$, and is 8 kpc (the solid curve in Fig. 6). The dotted curve 
in Fig. 6 indicates the expected variation in the observed angular size of an
object 8 kpc in diameter, as predicted by the model by Mao et al. (1998) 
(in this model, the galaxy size varies as  $\propto(1 + z)^{-1}$). Clearly, 
our data for the galaxies with $M(B) < -18$ are inconsistent with such a
strong evolution of their sizes. This conclusion agrees with the conclusion 
by Lilly et al. (1998) that the linear sizes of large and luminous spiral 
galaxies change little to $z\sim1$. A similar conclusion was reached by
Simard et al. (1999) and Takamiya (1999).

\begin{figure}[!ht]
\centerline{\psfig{file=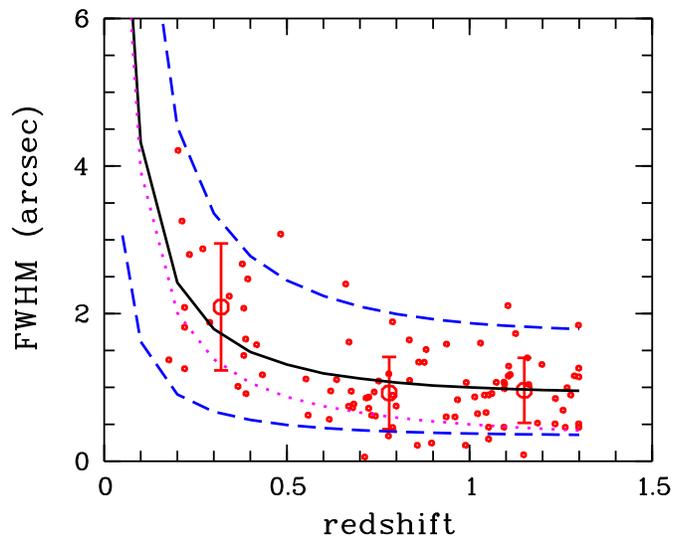,width=9cm,angle=-90,clip=}}
\caption{Galaxy angular size in $R_c$ versus redshift (see the text).}
\end{figure}

In the redshift range 0.3 to 0.5, where faint and luminous objects are 
represented almost equally, we compared the linear sizes for the galaxies 
with $M(B) > -18$ and $M(B) < -18$. The luminous ($M(B)\approx-19$) spiral 
and elliptical galaxies turned out to be a factor of about 1.5 to 2 larger 
than the fainter ($M(B)\approx-17$) galaxies. The expected change
in size for a galaxy with an exponential brightness distribution and a 
constant central surface brightness is $\Delta$log$D\propto$~0.2$\Delta$M . 
Therefore, when the absolute magnitude of a spiral galaxy changes by 2$^m$, 
its size will change by a factor of $\approx$2.5. The observed change
in galaxy size is smaller than this estimate, probably because the disks 
of many distant galaxies are poorly described by an exponential law 
(Reshetnikov et al. 2003).

The variation in the observed colors of extended objects with $z$ roughly 
corresponds to the expected variation for local galaxies at the 
corresponding $z$. As an example, Fig. 7a,b shows the galaxy positions
in the two-color diagrams for two redshift ranges. The solid line in each 
of the panels represents the color sequence for normal galaxies at 
$z = 0.2$ (a) and $z = 0.8$ (b), as constructed by Fukugita et al. (1995).
The colors of galaxies at $z = 0.7-0.9$ are in satisfactory agreement with 
the expected distribution in the two-color diagram (all of the objects in 
this $z$ range brighter than $M(B) = -18$); the agreement is poorer
for $z \approx 0.2$. However, the agreement becomes much better if we 
consider only the luminous galaxies with $M(B) < -18$ at $z \approx 0.2$ 
(the crosses in Fig. 7a).

\begin{figure*}[!ht]
\centerline{\psfig{file=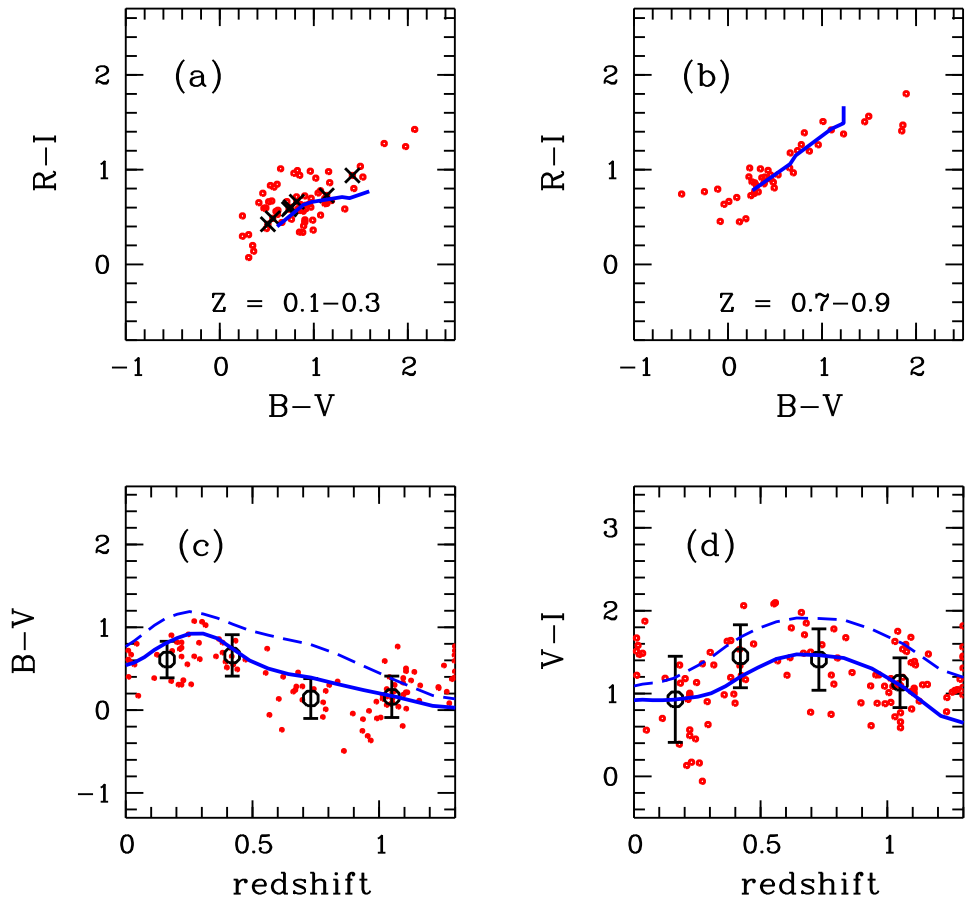,width=13cm,angle=-90,clip=}}
\caption{(a and b) Positions of the sample galaxies in the two-color diagrams 
for two redshift ranges. The solid lines indicate the corresponding color 
sequences for normal galaxies (Fukugita et al. 1995); the crosses in Fig. 7a 
mark luminous galaxies with $M(B) < -18$. (c and d) The evolution of the 
observed colors of spiral galaxies. The barred circles represent the mean
color indices in the redshift ranges 0--0.3, 0.3--0.6, 0.6--0.9, and 0.9--1.2. 
The heavy solid and dashed lines indicate the model dependences for the 
accretion model of disk galaxy formation and for the model of a single 
collapse (Westera et al. 2002), respectively. }
\end{figure*}

The color variations of Sa--Im spiral galaxies with $z$ are shown in a 
more detailed form in Figs. 7c and 7d. The solid lines in these figures 
represent the expected dependences for the accretion model of spiral
galaxy formation (Westera et al. 2002; Samland and Gerhard 2003). According 
to Westera et al. (2002), the galactic disk is formed inside a dark halo 
through ongoing external gas accretion. In this model, the star formation 
rate is a nonmonotonic function of time: it reaches its maximum at 
$z \approx 1$ and remains significant down to $z = 0$. The dashed lines 
indicate the dependences for the model of a single collapse. In this scenario, 
the disk has been formed through the contraction of a protogalactic cloud 
at $z > 1$. We see from Figs. 7c and 7d that, in general, the accretion
model satisfactorily describes the general pattern of variation in the 
observed colors of galaxies with $z$. The collapse model is in much poorer 
agreement with our data, predicting much redder colors at each $z$.

\subsection{The host galaxy of GRB\,000926}

What is the host galaxy of the GRB whose region is studied 
here?\footnote{An image of this galaxy and more information can be
found at SAO RAS http://www.sao.ru/hq/grb/host-obs.html.} 
The apparent magnitudes of the galaxy corrected for 
extinction in the Milky Way are $B = 26.01 \pm 0.17$, $V = 25.76 \pm 0.22$, 
$R_c = 25.88 \pm 0.28$, and $I_c = 24.65 \pm 0.40$ (Fatkhullin 2002); its
redshift is $z = 2.0379$ (Castro et al. 2003). Given the $k$-correction 
for an Sc spiral galaxy (Poggianti 1997), the absolute magnitude of the 
host galaxy is $M(B) = -19.1$. Consequently, it is a relatively faint, 
but not dwarf object. The galaxy mass was roughly estimated by Castro et al. 
(2003) to be $\approx10^{10}$M$_{\odot}$. This value yields an estimate 
of the mass-to-light ratio M/$L(B) \approx 1.5$ (in solar units), which is 
typical of objects of late morphological types with enhanced star
formation.

The observed colors of the host galaxy were compared with the colors of 
faint ($B > 25$) galaxies in this field by Fatkhullin (2002). The galaxy 
was found to be, on average, bluer than other faint extended objects.
This conclusion is also confirmed by a direct comparison of the observed 
colors for the host galaxy of GRB\,000926 with the colors of objects at 
$z \approx 2$ in our field. Note, however, that, because of observational
selection (see Fig. 5), the galaxies that we extracted at $z \approx 2$ 
are, on average, much more luminous than the host galaxy being discussed. 
A proper comparison of the colors requires studying the parameters of
galaxies at $z \approx 2$ with a luminosity comparable to that of the 
GRB\,000926 host galaxy. Since such objects are located near the magnitude 
limit of our field, this is very difficult to do using our data. However, 
certain conclusions can still be reached. That the colors of the host galaxy 
are bluer than the colors of galaxies at lower $z$ can be explained by the 
fact that the specific star formation rate in the Universe increases with
redshift (see, e.g., Madau et al. 1998); hence, bluer colors are expected, 
on average, for the galaxies (see, e.g., Fig. 3 from the paper by Rudnik 
et al. 2003). Therefore, the host galaxy may well be a typical spiral
galaxy for an epoch of $z \approx 2$. This explanation is consistent with 
the conclusion previously reached by several authors that the host 
galaxies of GRBs are normal blue galaxies similar in properties to galaxies
at the same redshifts (Sokolov et al. 2001; Le Floc'h et al. 2003; Djorgovski 
et al. 2003).

\section{Conclusions}

We have presented the results of our detailed photometric study of a small 
(3.6$'$$\times$3$'$) region in the field of GRB\,000926 (Fig. 1). 
The observations were carried out on good photometric nights, which allowed
us to reach a limiting magnitude of 26.6 in the $B$ band under fairly 
stringent conditions.

In the field under study, we extracted 264 extended objects and 21 starlike 
objects. The number of presumed stars agrees with the expected number in
Bezanson's Milky Way model. The stellar colors are typical of main sequence 
dwarfs (Fig. 2).

The differential galaxy counts in the field to $B \approx 25$ are in good 
(to within 20\%) agreement with previous CCD surveys of several deep fields 
(Fig. 3). We have confirmed the existence of a general tendency in the
observed colors of the field galaxies with their apparent magnitude (Fig. 4).

An analysis of the images of luminous ($M(B) < -18$) galaxies has led us 
to conclude that there is no strong evolution of their linear sizes at 
$z \leq 1$ (Fig. 6).

In general, the color variations of spiral galaxies with $z$ agree with 
the predictions of the accretion model, in which galactic disks are formed 
within dark halos through long-term external gas accretion (Fig. 7).

One of the important results of our study is that in the cases where we 
can compare our data with previously published data for other (often deeper 
and larger) fields, they are in good agreement. This implies that 
investigating relatively small deep fields is a quite efficacious method 
of studying the evolution of galaxies.

\bigskip
\section*{Acknowledgments}
{\it This study was supported by the Federal Program ``Astronomy'' 
(project no. 40.022.1.1.1101) and the Russian Foundation for Basic
Research (project nos. 03-02-17152 and 01-02-17106.}

\section*{REFERENCES}

\indent

S. Arnouts, B. Vandame, C. Benoist, et al., Astron.
Astrophys. 379, 740 (2001).

W. Baum, IAU Symp. 15: Problems of Extragalactic
Research (Macmillan, New York, 1963), p. 390.

E. Bertin and S. Arnouts, Astron. Astrophys. 117,
393 (1996).

M. Bessel, Publ. Astron. Soc. Pac. 102, 1181 (1990).

M. Bolzonella, J.-M. Miralles, and R. Pello, Astron.
Astrophys. 363, 476(2000).

D. Calzetti, L. Armus, R. C. Bohlin, et al., Astrophys.
J. 533, 682 (2000).

S. Castro, T. Galama, F. Harrison, et al., Astrophys.
J. 586, 128 (2003).

S. G. Djorgovski, S. R. Kulkarni, D. A. Frail, et al.,
Proc. SPIE 4834, 238 (2003).

T. Fatkhullin, Bull. Spec. Astrophys. Obs. 53, 5 (2002).

T. A. Fatkhullin, Candidate's Dissertation (SAO RAS,
Nizhnii Arkhyz, 2003); http://www.sao.ru/hq/grb/team/timur/timur.html.

H. C. Ferguson, M. Dickinson, and R. Williams, Ann.
Rev. Astron. Astrophys. 38, 667 (2000).

M. Fukugita, K. Shimasaku, and T. Ichikawa, Publ.
Astron. Soc. Pac. 107, 945 (1995).

J. P. Gardner, Publ. Astron. Soc. Pac. 110, 291
(1998).

M. Giavalisco et al. (the GOODS team), astro-ph/0309105 (2003).

J. Heidt, I. Appenzeller, A. Gabasch, et al., Astron.
Astrophys. 398, 49 (2003).

A. U. Landolt, Astron. J. 104, 340 (1992).

O. Le Fevre, Y. Mellier, H. J. McCracken, et al.,
astro-ph/0306252 (2003).

E. Le Floc'h, P.-A. Duc, I. Mirabel, et al., Astron.
Astrophys. 400, 499 (2003).

S. Lilly, D. Schade, R. Ellis, et al., Astrophys. J. 500,
75 (1998).

P. Madau, L. Pozzetti, and M. Dickinson, Astrophys.
J. 498, 106(1998).

T. Maihara, F. Iwamuro, H. Tanabe, et al., Publ. Astron. Soc. J. 
53, 25 (2001).

Sh. Mao, H. J. Mo, and S. D. M. White, Mon. Not.
R. Astron. Soc. 297, L71 (1998).

H. J. McCracken, M. Radovich, E. Bertin, et al.,
Astron. Astrophys. 410, 17 (2003).

N. Metcalfe, T. Shanks, A. Campos, et al., Mon. Not.
R. Astron. Soc. 323, 795 (2001).

B. M. Poggianti, Astron. Astrophys., Suppl. Ser. 122,
399 (1997).

E. Ramirez-Ruiz, E. Fenimore, and N. Trentham,
AIP Conf. Proc. 555, 457 (2001).

V. P. Reshetnikov, R.-J. Dettmar, and F. Combes,
Astron. Astrophys. 399, 879 (2003).

A. C. Robin, C. Reyle, and M. Creze, Astron. Astrophys. 359, 
103 (2000).

G. Rudnik, H.-W. Rix, M. Franx, et al., astro-ph/0307149 (2003).

M. Samland and O. E. Gerhard, Astron. Astrophys.
399, 961 (2003).

L. Simard, D. C. Koo, S. M. Faber, et al., Astrophys.
J. 519, 563 (1999).

V. Sokolov, T. Fatkhullin, A. Castro-Tirado, et al.,
Astron. Astrophys. 372, 438 (2001).

L. S. Sparke and J. S. Gallagher III, Galaxies in the
Universe: An Introduction (Cambridge Univ. Press,
Cambridge, 2000).

A. Szalay, A. Connolly, and G. Szokoly, Astron. J.
177, 68 (1999).

M. Takamiya, Astrophys. J., Suppl. Ser. 122, 109
(1999).

N. Trentham, E. Ramirez-Ruiz, and A. Blain, Mon.
Not. R. Astron. Soc. 334, 983 (2002).

D. Schlegel, D. Finkbeiner, and M. Davis, Astrophys.
J. 500, 525 (1998).

P. Westera, M. Samland, R. Buser, and O. E. Gerhard,
Astron. Astrophys. 389, 761 (2002).

\end{document}